# Temperature dependence of upper critical fields and coherence lengths for optimally-doped YBa$_2$Cu$_3$O$_{7-\delta}$ thin films

E. V. Petrenko[1], L. V. Omelchenko[1], A. V. Terekhov[1], Yu. A. Kolesnichenko[1], K. Rogacki[2], D. M. Sergeyev[3], and A. L. Solovjov[1,2,4]

[1]*B. Verkin Institute for Low Temperature Physics and Engineering of the National Academy of Sciences of Ukraine Kharkiv 61103, Ukraine*
E-mail: petrenko@ilt.kharkov.ua

[2]*Institute for Low Temperatures and Structure Research, Polish Academy of Sciences, Wroclaw 50-950, Poland*

[3]*K. Zhubanov Aktobe Regional State University, Aktobe 030000, Kazakhstan*

[4]*The faculty of physics, V. N. Karazin Kharkiv National University, Kharkiv 61022, Ukraine*

We report the comprehensive comparative analysis of the upper critical magnetic fields $\mu_0 H_{c2}(0)$ obtained within Ginzburg–Landau (GL) and Werthamer–Helfand–Hohenberg (WHH) theories for optimally-doped YBa$_2$Cu$_3$O$_{7-\delta}$ thin films. For different orientations of the magnetic field, our calculations give 638 and 153 T for $\mu_0 H_{c2}(0)$, $H \parallel ab$ and $\mu_0 H_{c2}(0)$, $H \parallel c$, respectively, when using $H_{c2}(0)$. For the first time, the temperature dependences of coherence lengths $\xi_{ab}(T)$ and $\xi_c(T)$ within proposed theories were determined using 50 and 90% criteria of the normal state resistivity value $\rho_N$. The GL (0.9$\rho_N$) approach gives $\xi_{ab}(0) = 11.8$ Å and $\xi_c(0) = 3.0$ Å which are in a good agreement with literature data. The implications of very short coherence lengths in HTSCs are discussed.

## 1. Introduction

The development of a rigorous theory capable of fully describing high-temperature superconductors (HTSCs) is still the most urgent task of modern solid state physics. Among HTSCs with a superconducting (SC) transition temperature $T_c$ exceeding the boiling point of liquid nitrogen ($T_{BN} = 77.4$ K), one can distinguish a class of metal oxides with an active plane CuO$_2$ such as YBa$_2$Cu$_3$O$_{7-\delta}$ (YBCO), called cuprates. It is known that, in addition to high $T_c$, these type-II superconductors have a number of specific properties, such as $d$-wave superconducting (SC) pairing, low charge carrier density, especially in the pseudogap (PG) state, strong electron correlations and quasi-two-dimensionality resulting in noticeable anisotropy, according to lots of research [1–6].

A high value of $T_c$ in cuprates is likely due to the small size of Cooper pairs, which is determined by the short in-plane coherence length $\xi_{ab}(0)$. It is well established that, due to anisotropy, $\xi_{ab}$ is approximately an order of magnitude larger than the coherence length along the $c$ axis $\xi_c$. Interestingly, in cuprates, the exact temperature dependence of both $\xi_{ab}$ and $\xi_c$ above and, more surprisingly, below $T_c$ has not yet been fully elucidated. To determine $\xi_{ab}(T)$ and $\xi_c(T)$ it is necessary to obtain the temperature dependences of the upper critical magnetic field $\mu_0 H_{c2}(T)$, preferably in a wide temperature range and in a magnetic field applied parallel to either the $ab$ plane or the $c$ axis, since the coherence lengths and the upper critical field are mutually correlated.

The in-plane electrical resistivity measurements relatively easy solve this problem when under- and/or over-doped cuprates are exposed to strong magnetic field generated, for example, within the 45-T hybrid direct current (dc) magnet and the 60- and 85-T pulsed magnets at the National High Magnetic Field Laboratory in Tallahassee, Florida and Los Alamos, New Mexico [7, 8]. However, in the case of optimally doped samples, much stronger magnetic fields are required [8]. Creating really strong magnetic fields (more than 140 T) becomes a much more difficult task, as it can lead to the inevitable destruction of not only the sample [9] but also the magnet itself [10]. Nevertheless, in [10] it is reported that pulsed magnetic fields of about 400 T were successfully achieved, which made it possible to try to construct the phase diagram of the upper critical field for an optimally doped YBCO single crystal in fields parallel and perpendicular to the $CuO_2$ planes. Of particular note is the pioneering work of Grissonnanche *et al.* [8], where combined measurements of thermal conductivity and electrical resistivity allowed to obtain the upper critical field $H_{c2}$ of the cuprate superconductor YBCO as a function of hole concentration (doping) $p$.

Upper critical field $\mu_0 H_{c2}(T)$ is a fundamental measure of the strength of superconductivity. However, despite the recent breakthrough in achieving extremely strong magnetic fields, there is still no consensus on the exact value of $\mu_0 H_{c2}(0)$ for optimally doped YBCO samples due to the complexity of the experimental procedures. All this leads to a large error at low temperatures near 0 K, especially when measuring in pulsed fields [8, 10]. Because of this obstacle, the chances of accurately determining the upper critical magnetic field in other cuprate HTSCs with even higher $T_c$ are also greatly reduced. For example, for $HgBa_2Ca_2Cu_3O_{8+\delta}$ (Hg-1223) superconductor, in which the SC transition temperature is about 134 K [11], the value of $\mu_0 H_{c2}(0)$ has not yet been estimated.

The question arises: which of the two theories mentioned above is capable of predicting the most adequate $\mu_0 H_{c2}(T)$ and $\mu_0 H_{c2}(0)$ value for optimally doped YBCO samples, if the available experimental equipment makes it possible to achieve a magnetic field of up to 14 T at best? In this paper, we report our attempts to answer this question by comparing our results with literature data. Having measured the SC transition curves of the sample in various magnetic fields and using the special software, we calculated the upper critical fields $\mu_0 H_{c2}(0)$ of the optimally doped YBCO film both within the Ginzburg–Landau (GL) [12, 13] and Wertamer–Helfand–Hochenberg (WHH) [14] theories. As a result, it was possible to reconstruct and plot complete curves $\mu_0 H_{c2}(T)$. The calculations were carried out for orientations of the external magnetic field parallel to both the *ab* plane and the *c* axis. We note a significant difference between the results obtained depending on the theory under consideration. The results obtained made it possible for the first time to calculate the temperature dependences of the coherence lengths in the *ab* plane $\xi_{ab}$ and along the *c* axis $\xi_c$. The effect of a small coherence length on the behavior of Cooper pairs is discussed in detail.

## 2. Experiment

Epitaxial YBCO films were deposited at a temperature $T = 770\ °C$ and an oxygen pressure of 3 mbar on $(LaAlO_3)_{0.3}(Sr_2TaAlO_6)_{0.7}$ substrates, as described in [15]. The thickness of the deposited films, $d \sim 100$ nm, was controlled by the deposition time of the corresponding targets. X-ray diffraction analyses showed that all samples are excellent films with the *c* axis perfectly oriented perpendicular to the $CuO_2$ planes. The films were then lithographically patterned and chemically etched into well-defined 2.35×1.24 mm Hall-bar structures. To make the contacts, the gold wires were glued to the pads of the structure using silver epoxy. Contact resistance below 1 Ω was obtained. The main measurements were performed using a fully computerized setup, including a Quantum Design Physical Property Measurement System (PPMS-9), using a drive current of ~100 μA at 19 Hz. The four-probe technique was used to measure the in-plane resistivity $\rho_{ab}(T) = \rho(T)$. During cooling, a constant magnetic field from zero to 9 T was sequentially applied to the sample to measure superconducting transitions.

## 3. Results and discussion

### 3.1. Resistivity

Figure 1 shows the temperature dependence of the resistivity $\rho(T)$ of the $YBa_2Cu_3O_{7-\delta}$ (YBCO) film in the range of 80–300 K in the absence of an external magnetic field. The dependence $\rho(T)$ shows a sharp decrease in the electrical resistivity below $T_c^{onset}$ and its disappearance at $T_c \sim 88$ K, which is typical for the transition to the SC state in well-structured YBCO thin films [16]. Figures 2(a) and 2(b) represent the temperature and field dependences of ρ in units of ($\rho/\rho_N$) for both $H \parallel ab$ and $H \parallel c$. Here $\rho_N$ is the resistivity of the normal state directly above the SC transition. It is noteworthy that the field is always applied perpendicular to the direction of the measuring current ($H \perp I$). As can be seen from Fig. 2, the magnetic field noticeably broadens the resistive transition, especially at *H* ∥ *c* [Fig. 2(b)], and reduces *T$_c$*, but, as usual, has almost no effect on the resistivity of the sample in the normal state [17, 18]. Unfortunately, for the case *H* = 9 T and *H* ∥ *ab*, contacts of the sample were damaged during the experiment and did not allow fixing the temperature dependence in this field.

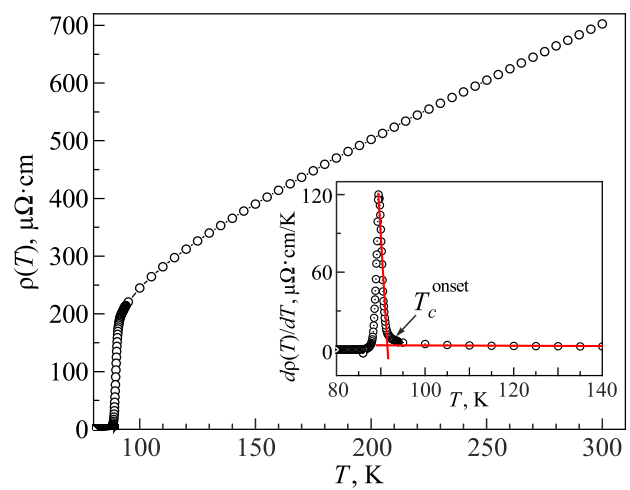

*Fig. 1.* Temperature dependence of the resistivity $\rho(T) = \rho_{ab}(T)$ of the $YBa_2Cu_3O_{7-\delta}$ film in the range 80–300 K in the absence of the external magnetic field. Inset represents a derivative of $\rho(T)$ at $H = 0$. The intersection of red straight lines defines $T_c^{onset}$.

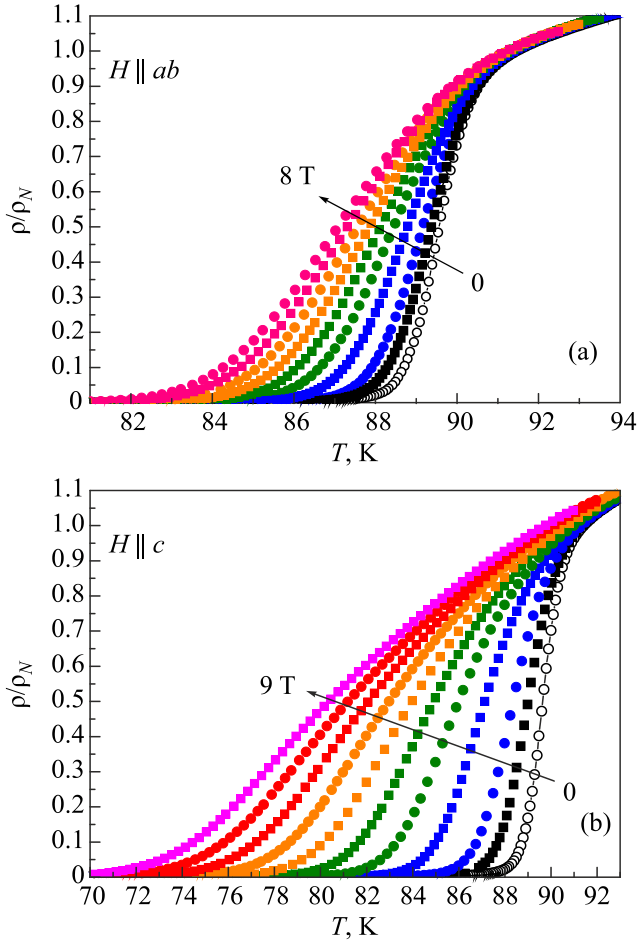

*Fig. 2.* (Color online) Temperature and magnetic field dependences of ρ(*T,H*) in units of (ρ/ρ$_N$) of the YBa$_2$Cu$_3$O$_{7-\delta}$ film. Figures 2(a) and 2(b) are obtained for the field orientations, parallel (*H* ∥ *ab*, μ$_0$*H* = 0, 0.5, 1, 2, 3, 4, 5, 6, 7, and 8 T) and perpendicular (*H* ∥ *c*, μ$_0$*H* = 0, 0.5, 1, 2, 3, 4, 5, 6, 7, 8, and 9 T) to the *ab* plane, respectively.

Unlike our previous work [19], now more precise values of $T_c^{onset}$ were determined for both *H* ∥ *ab* (91.48 K) and *H* ∥ *c* (91.44 K). To do this, we calculated the derivative of ρ(*T*) at *H* = 0, as shown in the inset to Fig. 1. Interestingly, when magnetic field is applied, the temperature of the onset of the SC transition practically does not change, while *T$_c$* (ρ = 0) noticeably decreases towards lower temperatures. It is worth emphasizing that we do not observe any steps at all SC transitions, which indicates the good quality of the sample, its homogeneity, and the absence of additional phases and inclusions.

### 3.2. Upper critical field

To calculate the temperature dependences of upper critical fields μ$_0$*H$_{c2}$*(*T*) for *H* ∥ *ab* and *H* ∥ *c* for the YBCO thin film under study, the data of Figs. 2(a) and 2(b) were used, respectively. To compare the results with those of many other authors [8–10, 20, 21], we considered ρ(*H*) data that meet 90 and 50% criterion of the normal state resistivity ρ$_N$ [20–22]. Recall that ρ$_N$ is the electrical resistivity of the sample at $T \geq T_c^{onset}$.

The red dashed curves in Figs. 3 and 4 are plotted according to the calculated data in the framework of the Werthamer–Helfand–Hohenberg (WHH) [14] theory which reads:

$$\ln\frac{1}{t} = \left(\frac{1}{2} + \frac{i\lambda_{so}}{4\gamma}\right)\psi\left(\frac{1}{2} + \frac{\bar{h} + \frac{1}{2}\lambda_{so} + i\gamma}{2t}\right)$$

$$+ \left(\frac{1}{2} - \frac{i\lambda_{so}}{4\gamma}\right)\psi\left(\frac{1}{2} + \frac{\bar{h} + \frac{1}{2}\lambda_{so} - i\gamma}{2t}\right) - \psi\left(\frac{1}{2}\right), \quad (1)$$

where ψ is the digamma function, and

$$\gamma \equiv \left[\left(\alpha\bar{h}\right)^2 - \left(\frac{1}{2}\lambda_{so}\right)^2\right]^{\frac{1}{2}}, \quad (2)$$

$$h^* = -\frac{H_{c2}}{(dH_{c2}/dt)|_{t=1}} = (\pi^2/4)\bar{h}, \quad (3)$$

which defines *h*\*, and *t* = *T/T$_c$*.

Formulas (1)–(3) written in the Mathcad 12.0 environment were used for the analysis, taking the Maki parameter α and the spin-orbit scattering parameter λ$_{so}$ as fitting parameters. Recall that α describes a relative contribution of the spin-paramagnetic (electron spins rotate in Cooper pairs under the influence of a magnetic field) effects and λ$_{so}$ describes the spin-orbit scattering (appear due to the Lorentz force acting on paired electrons) [14]. However, the best fit for our film was found at α = 0 and λ$_{so}$ = 0 (Figs. 3 and 4). This leads to a somewhat simplified WHH equation to estimate μ$_0$*H$_{c2}$*(*T*) without spin paramagnetism and spin-orbit interaction (Refs. 20, 22 and references therein). As can be seen from the insets to the figures, as the field increases, the experimental data deviate upward from the theoretical dependence constructed within the framework of the WHH theory both for *H* ∥ *c* and especially for *H* ∥ *ab*. This may indicate a strong coupling of electrons in a Cooper pair, *d*-wave pairing, and, possibly, the existence of unconventional superconductivity [23, 24]. Nevertheless, the experimental data provide reliable *dH$_{c2}$/dT* values used in the calculations. As mentioned above, the Maki parameter α = 0 and, therefore, μ$_0 H_{c2}^{orb}(0) = $ μ$_0 H_{c2}(0)$ and the spin paramagnetic effects are insignificant and appear only when α > 0.

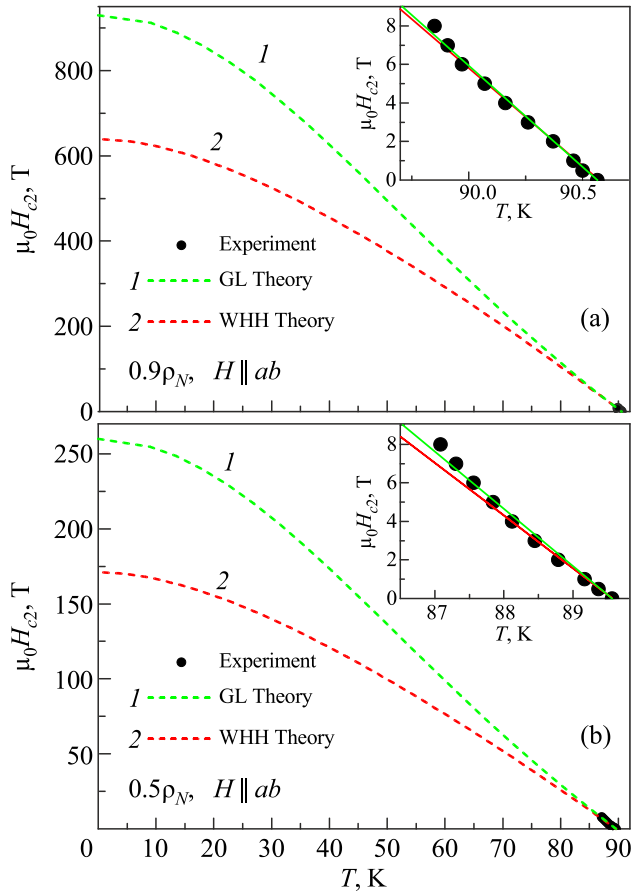

*Fig. 3.* (Color online) Temperature dependences of the upper critical magnetic fields $\mu_0 H_{c2}(T) \parallel ab$ determined at $\rho = 0.9\rho_N$ (a) and $\rho = 0.5\rho_N$ (b).

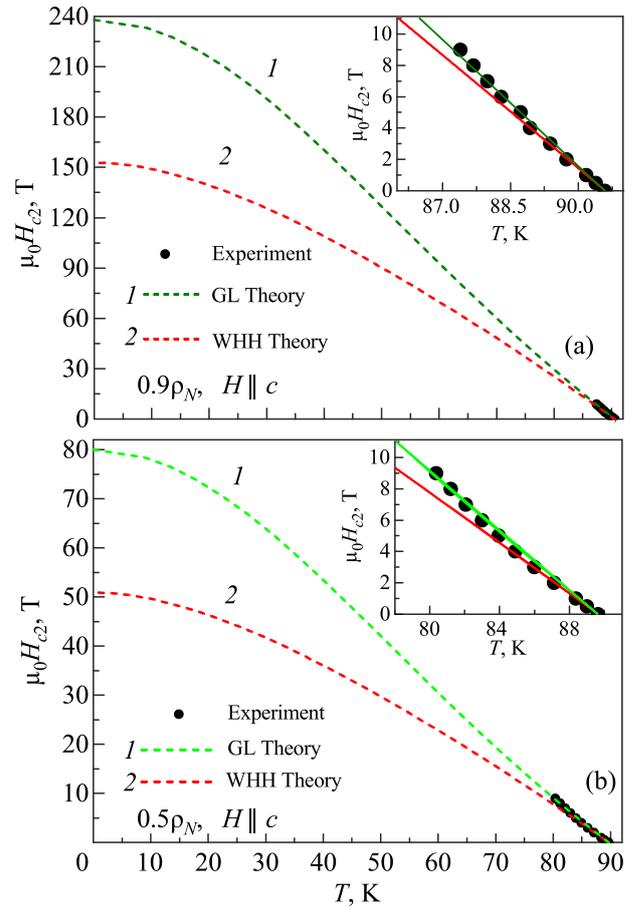

*Fig. 4.* (Color online) Temperature dependences of the upper critical magnetic fields $\mu_0 H_{c2}(T) \parallel c$ determined at $\rho = 0.9\rho_N$ (a) and $\rho = 0.5\rho_N$ (b).

Another approach to the problem is to use the formula from the Ginzburg–Landau (GL) theory [12, 13] to estimate the critical fields [20, 21]:

$$H_{c2}(T) = \frac{\Phi_0}{2\pi\xi^2(T)} = H_{c2}(0)\frac{(1-t^2)}{(1+t^2)}, \quad (4)$$

where, as before, $t = T/T_c$ and $\Phi_0 = 2.068 \cdot 10^{-15}$ T·m² is the magnetic flux quantum.

Due to the simplicity of the formulas in this case, the calculations were carried out directly in the Origin environment [23]. As can be seen from Figs. 3 and 4, the approximation by the GL theory describes the experimental data $\mu_0 H_{c2}(T)$ noticeably better for all field directions. At the same time, $\mu_0 H_{c2}(0)$ in this case is noticeably higher than when fitted within the WHH theory. Previously, such a feature has already been observed both in HTSCs and in magnetic superconductors, and it was noted that the values of $\mu_0 H_{c2}(0)$ for the GL theory often exceed the values of $\mu_0 H_{c2}(0)$ obtained from the WHH theory [21]. Table 1 shows the values of $\mu_0 H_{c2}(0)$ obtained as a result of calculations within the framework of both theories. The results obtained fully confirm the above conclusions.

From the analysis of the literature (Table 2), it can be seen that our data correlate with the results of other groups. A particularly good correlation is observed for the $0.5\rho_N$ criterion within both the GL and WHH theories, which give quite reasonable values of $\mu_0 H_{c2}(0)$ in both field directions. As can also be seen from Table 2, a significant role in the value of $\mu_0 H_{c2}(0)$ is played by the structure of the sample. Since it is believed that polycrystalline samples have the largest set of defects, the observed values of $\mu_0 H_{c2}(0)$ are the largest. For that reason, the smallest $\mu_0 H_{c2}(0)$ have single crystals [8, 10] and thin films [9] with minimal defects. Unfortunately, when measuring in pulsed magnetic fields, there is still a large error at low temperatures and especially near 0 K, which was rightly pointed out by the authors [9, 10]. In addition, there is an evident lack of results for $\mu_0H_{c2}(0)$ when $H \parallel ab$.

Table 1. Upper critical magnetic fields values (calculated within the framework of WHH and GL theories) for $H \parallel c$ and $H \parallel ab$

| Criterion | $\mu_0 H_{c2}(0)$, T; $H \parallel c$ ($T_c^{onset} = 91.44$ K) | | $\mu_0 H_{c2}(0)$, T; $H \parallel ab$ ($T_c^{onset} = 91.48$ K) | |
|---|---|---|---|---|
| | GL theory | WHH theory | GL theory | WHH theory |
| $0.9\rho_N$ | 238 | 153 | 930 | 638 |
| $0.5\rho_N$ | 80 | 51 | 260 | 171 |

Table 2. Literary data for $\mu_0H_{c2}(0)$ of optimally doped YBCO samples

| Refs. | YBCO sample | $T_c^{onset}$, K | $\mu_0H_{c2}(0)$, T; $H \parallel c$ | $\mu_0H_{c2}(0)$, T; $H \parallel ab$ | Type of fields |
|---|---|---|---|---|---|
| [20] | polycrystalline | 91 | WHH($0.9\rho_N$): 70 | – | Static |
| [21] | polycrystalline | 91 | GL($0.9\rho_N$): 563<br>GL($0.5\rho_N$): 70<br>WHH($0.9\rho_N$): 394<br>WHH($0.5\rho_N$): 52 | – | Static |
| [8] | single crystal | 91 | WHH:70 | – | Pulsed |
| [9] | film | 91 | WHH:70 | – | Pulsed |
| [10] | single crystal | ~90 | WHH:128 | WHH:240 | Pulsed |

The red and green dashed curves represent the results of calculations within the WHH and Ginzburg–Landau theories, respectively. The insets show the same dependences in the vicinity of $T_c^{onset}$, indicated by solid lines, and with larger experimental points.

The red and green dashed curves represent the results of calculations within the WHH and Ginzburg–Landau theories, respectively. The insets show the same dependences in the vicinity of $T_c^{onset}$, indicated by solid lines, and with larger experimental points.

### 3.3. Coherence lengths

Having determined the temperature dependences of the upper critical fields for various orientations of the magnetic field, one can calculate the corresponding dependences of the coherence lengths in the entire temperature range of interest using the following equations [25]:

$$\xi_{ab}(T) = \sqrt{\frac{\Phi_0}{2\pi\mu_0 H_{c2}(T)}}, \quad H \parallel c, \quad (5)$$

$$\xi_c(T) = \frac{\Phi_0}{2\pi\mu_0 H_{c2}(T)\xi_{ab}(T)}, \quad H \parallel ab. \quad (6)$$

The results are represented in Fig. 5, using in Eqs. (5) and (6) $\mu_0H_{c2}(T)$ found above for all four cases. The figure shows that, near $T_c$, the GL and WHH curves tend to coincide. At lower temperatures, the curves diverge, which is more pronounced when using the $0.5\rho_N$ criterion.

The calculated values of coherence lengths $\xi_{ab}(0)$ and $\xi_c(0)$ and the anisotropy factor $\Gamma = \xi_{ab}(0)/\xi_c(0)$ are listed in Table 3. It should be emphasized that our results are in good agreement with the literature data, which are in the range $\xi_{ab}(0) = (11.2–16)$ Å and $\xi_c(0) = (1.4–3.8)$ Å, respectively [16–18, 26, 27]. As can be seen from Table 3, the most correct values of the corresponding coherence lengths ($\xi_{ab}(0) = 11.8$ Å and $\xi_c(0) = 3.0$ Å) are obtained within the GL theory using the $0.9\rho_N$ criterion, where the average anisotropy factor is roughly equal to 4.0. Nevertheless, $\xi_{ab}(0) = 14.7$ Å and $\xi_c(0) = 3.5$ Å obtained within WHH theory at $0.9\rho_N$ criterion and with $\Gamma = 4.2$ are also reasonable.

As can be seen from Table 2, only Sekitani *et al*. [10] was able to measure $\mu_0H_{c2}(0)$ in all directions of the mag-netic field while studying the YBCO single crystal. Interestingly, our $\mu_0 H_{c2}(0) = 260$ T, $H \parallel ab$ (Table 1) calculated by the $0.5\rho_N$ criterion within the framework of the GL theory, is quite close to $\mu_0 H_{c2}(0) = 240$ T, $H \parallel ab$ found in [10], which indicates the correctness of our approach. Somewhat surprisingly, the authors [10] did not calculate either $\xi_{ab}(0)$ and $\xi_c(0)$, or the temperature dependences of $\xi_{ab}$ and $\xi_c$. Taking their results shown in Table 2, we calculated both $\xi_{ab}(0)$ and $\xi_c(0)$ and found them to be 16.0 and 8.6 Å, of which the latter is noticeably larger than the generally accepted results. Moreover, the anisotropy factor appeared to be roughly equal 1.9, which is rather small. Most likely, this indicates that the use of very strong pulsed magnetic fields can lead to the appearance of additional defects in the sample under study. However, further research is needed to elucidate this issue.

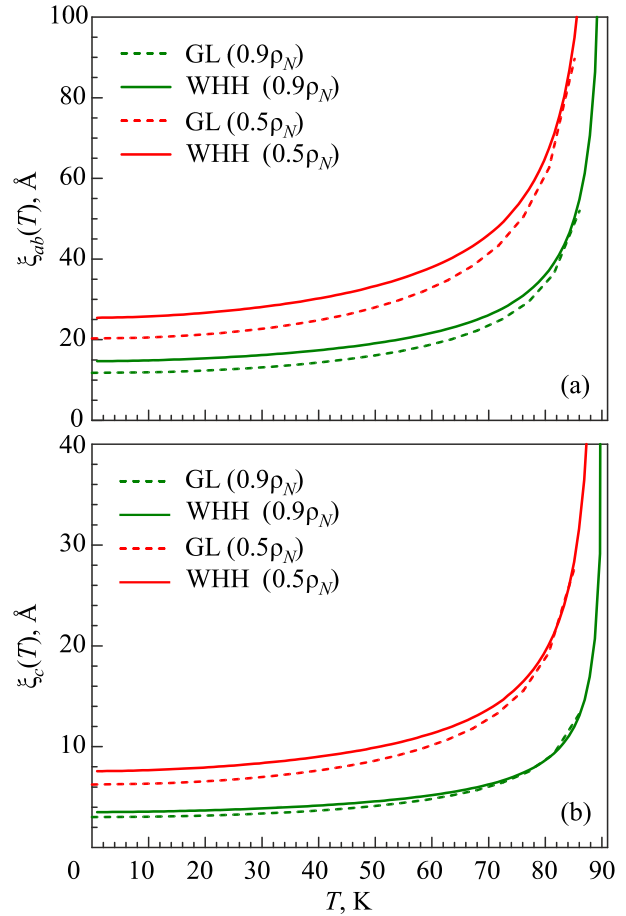

*Fig. 5.* (Color online) Temperature dependences of $\xi_{ab}(T)$ [Fig. 5(a)] and $\xi_c(T)$ [Fig. 5(b)] coherence lengths calculated from dependences $\mu_0H_{c2}(T)$, $H \parallel ab$ and $\mu_0H_{c2}(T)$, $H \parallel c$ (Figs. 3 and 4) using GL (dashed curves) and WHH (solid curves) theories. The criteria $0.9\rho_N$ (green curves) and $0.5\rho_N$ (red curves) are taken into account.

Table 3. Calculated coherence lengths $\xi_{ab}(0)$ and $\xi_c(0)$ and anisotropy factor $\Gamma = \xi_{ab}(0)/\xi_c(0)$ depending on the theory (GL or WHH) and criterion ($0.9\rho_N$ or $0.5\rho_N$)

| Criterion | GL theory | | | WHH theory | | |
|---|---|---|---|---|---|---|
| | $\xi_{ab}(0)$, Å | $\xi_c(0)$, Å | $\Gamma$ | $\xi_{ab}(0)$, Å | $\xi_c(0)$, Å | $\Gamma$ |
| $0.9\rho_N$ | 11.8 | 3.0 | 3.93 | 14.7 | 3.5 | 4.20 |
| $0.5\rho_N$ | 20.3 | 6.2 | 3.27 | 25.4 | 7.6 | 3.34 |

### 3.4. Implications of very short coherence lengths in HTSCs

Taking into account the results obtained (Figs. 5(a) and 5(b), and Table 3), we can consider the question: why is the second critical magnetic field in YBCO and other HTSCs so large? Most likely, this is due to the extremely short coherence length (Table 3), that is, the extremely small size of superconducting pairs in HTSCs at low temperatures. At first glance, this conclusion follows directly from the simple formula $H_{c2}(T) = \Phi_0 / 2\pi\xi(T)^2$ [17]. However, it is very likely that in HTSCs it is necessary to take into account the specific properties of Cooper pairs (CPs), which must change with temperature. Thus, a small coherence length should lead to an extremely high binding energy $\varepsilon_b \sim 1/\xi^2$ inside the pair [28, 29] both at $T$ near the PG opening temperature $T^*$ and at very low temperatures. This suggests that the unexpectedly high value of $\mu_0 H_{c2}(0)$ observed for YBCO systems can be explained in terms of strongly bound bosons [30, 31]. However, surprisingly, in HTSCs the evolution of local pairs in the PG state at $T^* > T > T_c$ ([30–33] and references therein), has been studied much better than the evolution of the CPs below $T_c$.

According to the relevant theories [1–4, 30], the local pairs above $T_c$ appear in HTSCs just below $T^*$ in the form of so-called strongly bound bosons (SBBs), which obey the theory of Bose–Einstein condensation (BEC) [5, 6, 30]. Since $\xi_{ab}(0) = \xi_{ab}(T^*)$ is roughly 13 Å (Table 3), the size of the SBB is extremely small, but the binding energy within the pair, $\varepsilon_b \sim 1/\xi_{ab}^2$, is extremely high [28, 29, 31]. In addition, the amount of SBBs at high temperatures is relatively small. As a result, SBBs do not interact with each other and, more importantly, are not destroyed by any reasonable external influence. Moreover, since $\xi_{ab}(T) \gg \xi_c(T)$ the pairs are almost flat and confined within the $CuO_2$ planes [34].

It is believed that as the temperature decreases, $\xi_{ab}(T)$ and, accordingly, the pair size increases according to the same law as when the temperature rises below $T_c$, as shown in Fig. 5. At the next characteristic temperature above $T_c$, usually denoted as $T_{pair}$, the local pairs become large enough and begin interact with each other. In addition, below $T_{pair}$, they acquire the ability to be destroyed by thermal fluctuations, i.e., they gradually transform into fluctuating CPs obeying the BCS theory. Thus $T_{pair}$, which is about 133 K for YBCO and is independent of oxygen doping, is the BEC-BCS transition temperature [30, 31]. The transition appears as a maximum in the temperature dependence of PG [31, 35] or as a deviation of the spectral weight loss close to the Fermi level, $W_{EF}(T)$, measured by the ARPES method, from a linear dependence at high temperatures [33].

Near $T_c$ both $\xi_{ab}(T)$ and $\xi_c(T)$ become very large, and the HTSC goes into a relatively homogeneous state of three dimensional (3D) SC fluctuations, where local pairs behave in a good many way like SC Cooper pairs but without long-range correlation interaction [1–6, 30] (the so-called short-range phase correlations). This conclusion is confirmed by numerous measurements of fluctuation conductivity in various HTSCs (see [35–37] and references therein).

By analogy, we can assume that below $T_c$, the same evolution of SC pairs should be observed. Indeed, near $T_c$ the $\xi_{ab}(T)$ and $\xi_c(T)$ are very large, and it is assumed that the HTCS is in a 3D state. This conclusion is confirmed by the 3D temperature dependence of the critical current density [38] measured in well-structured YBCO thin films near $T_c$ [39]. The three-dimensional dependence changes rapidly, since both $\xi_{ab}(T)$ and $\xi_c(T)$ and, accordingly, the sizes of the SC pairs sharply decrease with decreasing $T$. Below about 45 K, $\xi_{ab}(T)$, $\xi_c(T)$ and, accordingly, the both sizes of the SC pairs become very small and, interestingly, almost do not change with temperature (refer to Fig.5). Consequently, the binding energy inside the pairs must again be very large, and it is very tempting to conclude that at low temperatures the SC pairs should behave like SBBs. If this is really the case, this could explain the very high critical magnetic field $\mu_0 H_{c2}(0)$ observed in HTSCs, since SBBs are very difficult to destroy with a small magnetic field, as mentioned above.

Surprisingly, as far as we know, there is a clear lack of both experimental and theoretical results that could shed light on the possible evolution of CPs in HTSCs with decreasing $T$ below $T_c$. A rather interesting result was obtained by Albrecht [40], who measured the critical current density $J_c(T)$ in well-structured YBCO thin films down to very low temperatures. A noticeable increase in $J_c(T)$ was observed below $T_{SBB} \approx 43$ K for all samples, which is about 45 K below $T_c \approx 88$ K. The result becomes reasonable if we remember that in YBCO $T_{pair} \sim 133$ K, that is, also about 45 K, but above $T_c$. Recall, that for $T \geq T_{pair}$ local pairs turn into SSBs.

Thus, the result obtained below $T_c$ indicates that, by analogy with $T_{pair}$, it is quite possible that CPs in HTSC are converted to SBBs when the temperature is lowered below $T_{SBB}$, which confirms the above conclusion. In turn, $T_{SBB}$ can most likely be considered as the SC analogue of $T_{pair}$. In addition, since $\xi_c(T)$ is very small, the CPs are also believed to be confined in the $CuO_2$ planes below the $T_{SBB}$. However, there is still no idea whether there is any correlation interaction between adjacent $CuO_2$ planes in HTSC at such low $T$, and also what is the nature of the pairing mechanism that allows CPs in the form of SBBs to survive when such strong magnetic fields are applied.

Another open question is: what happens to SC Cooper pairs in HTSC when $T$ increases and passes through $T_c$? In conventional superconductor, everything is clear: the SC gap decreases when $T$ approaches $T_c$ in accordance with the law of the BCS theory and drops to zero at $T_c$. Accordingly, the size of the CPs tends to infinity at $T_c$, which means that the pairs are destroyed. Thus, as expected, there are no CPs above $T_c$, except for a very narrow range of SC fluctuations (~0.2 K) observed in thin SC films ([16] and references therein).

A completely different picture is observed in HTSCs. As $T$ approaches $T_c$, the temperature dependence of SC gap deviates upward from the BCS theory and never decreases to zero, as is observed in both YBCO [41] and BiSCCO ($Bi_2Sr_2Ca_2Cu_3O_{10+\delta}$) [41, 42] cuprates. Moreover, ARPES measurements convincingly show that the SC gap gradually becomes PG, when $T$ increases and passes through $T_c$ [43, 44]. Thus, we can conclude that the SC Cooper pairs do not collapse at $T=T_c$, but only lose their long-range SC correlation. The presence of such pairs above $T_c$ is predicted by the theory [3, 30, 32] and confirmed in various experimental studies [41, 42, 45, 46], including numerous fluctuation conductivity measurements ([35, 36] and references therein). However, the clearest evidence in favour of this conclusion is provided by measurements of the excess current, $I_{exc}$, performed on Ag-BiSCCO point contacts with $T_c \approx 110$ K [47]. The excess current was observed over a wide temperature range from 77 K and, more importantly, up to 206 K. It is known that the excess current is due to the Andreev reflection effect [48], which is possible only in the presence of a condensate of CPs [49]. The corresponding theory says that experimental observation of Andreev reflection effects above $T_c$ provides unambiguous evidence for the existence of uncorrelated CPs in the pseudogap state of cuprates [50]. Thus, the experiment clearly confirmed the existence of uncorrelated CPs in HTSCs well above $T_c$. It should also be emphasized that the corresponding current-voltage characteristics pass through $T_c$ without any peculiarities. This fact should once again confirm the above conclusion that SC Cooper pairs do not collapse at $T = T_c$ and gradually become uncorrelated CPs above $T_c$. But, strictly speaking, the physics of this process is still very mysterious.

### Conclusions

We have performed the comprehensive analysis of the upper critical magnetic fields $\mu_0 H_{c2}(0)$ obtained within Ginzburg–Landau and Werthamer–Helfand–Hohenberg theories for optimally-doped $YBa_2Cu_3O_{7-\delta}$ thin films. The results obtained for different orientations of the magnetic field and calculated within both theories using the $0.5\rho_N$ criteria are in a good agreement with the literature data measured with pulsed magnets. From the found $\mu_0 H_{c2}(T)$, with the appropriate software, the temperature dependences of coherence lengths $\xi_{ab}(T)$ and $\xi_c(T)$ were derived. Interestingly, now using the $0.9\rho_N$ criterion gives more correct $\xi(T)$ values. It has been shown for the first time that below $T_{SBB} \sim 40$ K, both $\xi_{ab}(T)$ and $\xi_c(T)$ acquire the smallest values, which practically do not change with a further decrease in temperature towards $T = 0$ K (Fig. 5). The extremely short $\xi(T)$ means the very high binding energy $\varepsilon_b \sim 1/\xi^2$ inside the SC Cooper pairs. The fact allows us to conclude that below the $T_{SBB}$, conventional Cooper pairs are transformed into strongly bound bosons (SBB), as happens at $T = T_{pair}$ above $T_c$. SBBs are extremely tightly bound pairs that cannot be destroyed by a relatively weak magnetic field. We believed that this explains the extremely high upper critical magnetic field observed in cuprates. It has been shown that the extremely short coherence lengths in HTSCs are responsible for the unusual behavior of CPs, observed when the temperature passes through $T_c$. The implication of this conclusion is discussed in detail.

### Acknowledgments

We thank Yuri Naidyuk, Vladimir Gnezdilov, Vladimir Tarenkov, and Aleksandr D'yachenko for valuable comments and discussions. We acknowledge support from the National Academy of Sciences of Ukraine through Young Scientists Grant No. 1/N-2021 (E.V.P. and L.V.O.) and also from the Ministry of Innovative Development of the Republic of Uzbekistan through Grant No. Ф-ФА-2021-433 (A.L.S.) and the Science Committee of the Ministry of Education and Science of the Republic of Kazakhstan through Grant No. AP08052562 (E.V.P. and D.M.S.). A.L.S. also thanks the Division of Low Temperatures and Superconductivity, INTiBS Wroclaw, Poland, for their hospitality.

---

1. V. M. Loktev, R. M. Quick, and S. G. Sharapov, *Phase fluctuations and pseudogap phenomena*, *Phys. Rep.* **349**, 1 (2001).

2. S. S. Dash and D. Sénéchal, *Pseudogap transition within the superconducting phase in the three-band Hubbard model*, *Phys. Rev. B* **100**, 214509 (2019).

3. V. Mishra, U. Chatterjee, J. C. Campuzano, and M. R. Norman, *Effect of the pseudogap on the transition temperature in the cuprates and implications for its origin*, Nat. Phys. **10**, 357 (2014).
4. N. J. Robinson, P. D. Johnson, T. M. Rice, and A. M. Tsvelik, *Anomalies in the pseudogap phase of the cuprates: Competing ground states and the role of umklapp scattering*, Rep. Prog. Phys. **82**, 126501 (2019).
5. L. Taillefer, *Scattering and Pairing in Cuprate Superconductors*, Annu. Rev. Condens. Matter Phys. **1**, 51 (2010).
6. S. Badoux, W. Tabis, F. Laliberté, G. Grissonnanche, B. Vignolle, D. Vignolles, J. Béard, D. A. Bonn, W. N. Hardy, R. Liang, N. Doiron-Leyraud, Louis Taillefer, and Cyril Proust, *Change of carrier density at the pseudogap critical point of a cuprate superconductor*, Nature (London) **531**, 210 (2016).
7. C. Tarantini, A. Gurevich, J. Jaroszynski, F. Balakirev, E. Bellingeri, I. Pallecchi, C. Ferdeghini, B. Shen, H. H. Wen, and D. C. Larbalestier, *Significant enhancement of upper critical fields by doping and strain in iron-based superconductors*, Phys. Rev. B **84**, 184522 (2011).
8. G. Grissonnanche, O. Cyr-Choinière, F. Laliberté, S. Rene de Cotret, A. Juneau-Fecteau, S. Dufour-Beauséjour, M.-E. Delage, D. LeBoeuf, J. Chang, B. J. Ramshaw, D. A. Bonn, W. N. Hardy, R. Liang, S. Adachi, N. E. Hussey, B. Vignolle, C. Proust, M. Sutherland, S. Kramer, J.-H. Park, D. Graf, N. Doiron-Leyraud, and Louis Taillefer, *Direct measurement of the upper critical field in cuprate superconductors*, Nature Commun. **5**, 3280 (2014).
9. J. L. Smith, J. S. Brooks, C. M. Fowler, B. L. Freeman, J. D. Goettee, W. L. Hults, J. C. King, P. M. Mankiewich, E. I. De Obaldia, M. L. O'Malley, D. G. Rickel, and W. J. Skocpol, *Low-temperature critical field of YBCO*, J. Supercond. **7**, 269 (1994).
10. T. Sekitani, N. Miura, S. Ikeda, Y. H. Matsuda, and Y. Shiohara, *Upper critical field for optimally-doped* $YBa_2Cu_3O_{7-\delta}$, Physica B **346–347**, 319 (2004).
11. Y. Moriwaki, T. Sugano, C. Gasser, A. Fukuoka, K. Nakanishi, S. Adachi, and K. Tanabe, *Epitaxial* $HgBa_2Ca_2Cu_3O_y$ *films on* $SrTiO_3$ *substrates prepared by spray pyrolysis technique*, Appl. Phys. Lett. **69**, 3423 (1996).
12. V. L. Ginzburg and L. D. Landau, *On the theory of superconductivity*, Zh. Eksp. Teor. Fiz. **20**, 1064 (1950).
13. L. Wang, H. S. Lim, and C. K. Ong, *Upper critical fields and order parameters of layered superconductors in a continuous Ginzburg–Landau model*, Supercond. Sci. Technol. **14**, 754 (2001).
14. N. R. Werthamer, K. Helfand, and P. C. Hohenberg, *Temperature and purity dependence of the superconducting critical field*, $H_{c2}$. *III. Electron spin and spin-orbit effects*, Phys. Rev. **147**, 295 (1966).
15. P. Przyslupski, I. Komissarov, W. Paszkowicz, P. Dluzewski, R. Minikayev, and M. Sawicki, *Structure and magnetic characterization of* $La_{0.67}Sr_{0.33}MnO_3/YBa_2Cu_3O_7$ *superlattices*, J. Appl. Phys. **95**, 2906 (2004).
16. A. L. Solovjov and V. M. Dmitriev, *Fluctuation conductivity and pseudogap in YBCO high-temperature superconductors*, Fiz. Nizk. Temp. **35**, 227 (2009) [Low Temp. Phys. **35**, 169 (2009)].
17. B. Oh, K. Char, A. D. Kent, M. Naito, M. R. Beasley, T. H. Geballe, R. H. Hammond, A. Kapitulnik, and J. M. Graybeal, *Upper critical field, fluctuation conductivity, and dimensionality of* $YBa_2Cu_3O_{7-x}$, Phys. Rev. B **37**, 7861 (1988).
18. E. Nazarova, A. Zaleski, and K. Buchkov, *Doping dependence of irreversibility line in* $Y_{1-x}Ca_xBa_2Cu_3O_{7-\delta}$, Physica C **470**, 421 (2010).
19. E. V. Petrenko, L. V. Omelchenko, Yu. A. Kolesnichenko, N. V. Shytov, K. Rogacki, D. M. Sergeyev, and A. L. Solovjov, *Study of fluctuation conductivity in* $YBa_2Cu_3O_{7-\delta}$ *films in strong magnetic fields*, Fiz. Nizk. Temp. **47**, 1148 (2021) [Low Temp. Phys. **47**, 1050 (2021)].
20. P. Rani, A. Pal, and V. P. S. Awana, *High field magneto-transport study of* $YBa_2Cu_3O_7:Ag_x$ ($x = 0.00–0.20$), Physica C **497**, 19 (2014).
21. R. Sultana, P. Rani, A. K. Hafiz, R. Goyal, and V. P. S. Awana, *An inter comparison of the upper critical fields* ($H_{c2}$) *of different superconductors* — $YBa_2Cu_3O_7$, $MgB_2$, $NdFeAsO_{0.8}F_{0.2}$, $FeSe_{0.5}Te_{0.5}$, *and* $Nb_2PdS_5$, Lett. J. Sup. Novel. Mag. **29**, 1399 (2016).
22. E. Nazarova, N. Balchev, K. Nenkov, K. Buchkov, D. Kovacheva, A. Zahariev, and G. Fuchs, *Transport and pinning properties of Ag-doped* $FeSe_{0.94}$, Supercond. Sci. Technol. **28**, 025013 (2015).
23. A. V. Terekhov, I. V. Zolochevskii, L. A. Ischenko, A. N. Bludov, A. Zaleski, E. P. Khlybov, and S. A. Lachenkov, *Magnetic ordering and specific features of its coexistence with superconductivity in* $Dy_{0.6}Y_{0.4}Rh_{3.85}Ru_{0.15}B_4$, Fiz. Nizk. Temp. **45**, 1467 (2019) [Low Temp. Phys. **45**, 1241 (2019)].
24. G. Cheng, M. Tomczyk, S. Lu, Joshua P. Veazey, Mengchen Huang, Patrick Irvin, Sangwoo Ryu, Hyungwoo Lee, Chang-Beom Eom, C. Stephen Hellberg, and Jeremy Levy, *Electron pairing without superconductivity*, Nature Lett. **521**, 196 (2015).
25. M. Bristow, W. Knafo, P. Reiss, W. Meier, P. C. Canfield, S. J. Blundell, and A. I. Coldea, *Competing pairing interactions responsible for the large upper critical field in a stoichiometric iron-based superconductor* $CaKFe_4As_4$, Phys. Rev. B **101**, 134502 (2020).
26. Y. Matsuda, T. Hirai, S. Komiyama, T. Terashima, Y. Bando, K. Iijima, K. Yamamoto, and K. Hirata, *Magnetoresistance of c axis-oriented epitaxial* $YBa_2Cu_3O_{7-x}$ *filmls above* $T_c$, Phys. Rev. B **40**, 5176 (1989).
27. J. Sugawara, H. Iwasaki, N. Kobayashi, H. Yamane, and T. Hirai, *Fluctuation conductivity of a c axis-oriented* $Yba_2Cu_3O_y$ *film prepared by chemical vapor deposition*, Phys. Rev. B **46**, 14818 (1992).
28. J. R. Engelbrecht, A. Nazarenko, M. Randeria, and E. Dagotto, *Pseudogap above* $T_c$ *in a model with* $d_{x^2-y^2}$ *pairing*, Phys. Rev. B **57**, 13406 (1998).
29. R. Haussmann, *Properties of a Fermi liquid at the superfluid transition in the crossover region between BCS superconductivity and Bose–Einstein condensation*, Phys. Rev. B **49**, 12975 (1994).

30. Mohit Randeria, *Pre-pairing for condensation*, *Nature Phys.* **6**, 561 (2010).
31. A. L. Solovjov and V. M. Dmitriev, *Resistive studies of the pseudogap in YBCO films with consideration of the transition from BCS to Bose–Einstein condensation*, *Fiz. Nizk. Temp.* **32**, 139 (2006) [*Low Temp. Phys.* **32**, 99 (2006)].
32. O. Tchernyshyov, *Non-interacting Cooper pairs inside a pseudogap*, *Phys. Rev. B* **56**, 3372 (1997).
33. T. Kondo, Y. Hamaya, A. D. Palczewski, T. Takeuchi, J. S. Wen, Z. J. Xu, G. Gu, J. Schmalian, and A. Kaminski, *Disentangling Cooper-pair formation above the transition temperature from the pseudogap state in the cuprates*, *Nature Phys.* **6**, 21 (2011).
34. Y. B. Xie, *Superconducting fluctuations in the high-temperature superconductors: Theory of the dc resistivity in the normal state*, *Phys. Rev. B* **46**, 13997 (1992).
35. A. L. Solovjov, L. V. Omelchenko, V. B. Stepanov, R. V. Vovk, H.-U. Habermeier, H. Lochmajer, P. Przysłupski, and K. Rogacki, *Specific temperature dependence of pseudogap in* $YBa_2Cu_3O_{7-\delta}$ *nanolayers*, *Phys. Rev. B* **94**, 224505 (2016).
36. A. L. Solovjov, E. V. Petrenko, L. V. Omelchenko, R. V. Vovk, I. L. Goulatis, and A. Chroneos, *Effect of annealing on a pseudogap state in untwinned* $YBa_2Cu_3O_{7-\delta}$ *single crystals*, *Sci. Rep.* **9**, 9274 (2019).
37. M. S. Grbiĉ, M. Požek, D. Paar, V. Hinkov, M. Raichle, D. Haug, B. Keimer, N. Bariĉiĉ, and A. Dulĉiĉ, *Temperature range of superconducting fluctuations above* $T_c$ *in* $YBa_2Cu_3O_{7-\delta}$ *single crystals*, *Phys. Rev. B* **83**, 144508 (2011).
38. E. A. Pashitsky, V. I. Vakaryuk, S. M. Ryabchenko, and Yu. V. Fedotov, *Temperature dependence of the critical current in high-temperature superconductors with low-angle interfaces between crystalline blocks*, *Fiz. Nizk. Temp.* **27**, 131 (2001) [*Low Temp. Phys.* **27**, 96 (2001)].
39. A. L. Solovjov, V. M. Dmitriev, V. N. Svetlov, and V. B. Stepanov, *Fluctuation conductivity and critical currents in Y–Ba–Cu–O films*, *Fiz. Nizk. Temp.* **29**, 1281 (2003) [*Low Temp. Phys.* **29**, 973 (2003)].
40. J. Albrecht, *Temperature-dependent pinning of vortices in low-angle grain boundaries in* $YBa_2Cu_3O_{7-\delta}$, *Phys. Rev. B* **68**, 054508 (2003).
41. J. L. Tallon, F. Barber, J. G. Storey, and J. W. Loram, *Coexistence of the superconducting energy gap and pseudogap above and below the transition temperature of superconducting cuprates*, *Phys. Rev. B* **87**, 140508 (2013).
42. Y. Yamada, K. Anagawa, T. Shibauchi, T. Fujii, T. Watanabe, A. Matsuda, and M. Suzuki, *Interlayer tunneling spectroscopy and doping-dependent energy-gap structure of the trilayer superconductor* $Bi_2Sr_2Ca_2Cu_3O_{10+\delta}$, *Phys. Rev. B* **68**, 054533 (2003).
43. K. Nakayama, T. Sato, Y. Sekiba, K. Terashima, P. Richard, T. Takahashi, K. Kudo, N. Okumura, T. Sasaki, and N. Kobayashi, *Evolution of a Pairing-Induced Pseudogap from the Superconducting Gap of* $(Bi,Pb)_2Sr_2CuO_6$, *Phys. Rev. Lett.* **102**, 227006 (2009).
44. T. Kondo, A. D. Palczewski, Y. Hamaya, T. Takeuchi, J. S. Wen, Z. J. Xu, G. Gu, and A. Kaminski, *Phys. Rev. Lett.* **111**, 157003 (2013).
45. K. Kawabata, S. Tsukui, Y. Shono, O. Michikami, H. Sasakura, H. Sasakura, K. Yoshiara, and T. Yotsuya, *Detection of a coherent boson current in the normal state of a high-temperature superconductor* $YBa_2Cu_3O_y$ *film patterned to micrometer-sized rings*, *Phys. Rev. B* **58**, 2458 (1998).
46. J. Corson, R. Mallozzi, J. Orenstein, J. N. Eckstein, and I. Bozovic, *Vanishing of phase coherence in underdoped* $Bi_2Sr_2CaCu_2O_{8+\delta}$, *Nature (London)* **398**, 221 (1999).
47. A. I. D'yachenko, V. Yu. Tarenkov, S. L. Sidorov, V. N. Varyukhin, and A. L. Solovjov, *The problem of pseudogap and excess current in* $Bi2223$–$Ag$ *contact at* $T > T_c$, *Fiz. Nizk. Temp.* **39**, 416 (2013) [*Low Temp. Phys.* **39**, 323 (2013)].
48. A. F. Andreev, *Thermal conductivity of the intermediate state of superconductors*, *Sov. JETP* **46**, 1823 (1964).
49. G. Deutscher, *Nature (London)* **397**, 410 (1999).
50. H.-Y. Choi, Y. Bang, and D. K. Campbell, *Andreev reflections in the pseudogap state of cuprate superconductors*, *Phys. Rev. B* **61**, 9748 (2000).

___

E. V. Petrenko, L. V. Omelchenko, A. V. Terekhov, Yu. A. Kolesnichenko, K. Rogacki, D. M. Sergeyev, and A. L. Solovjov

Проведено всесторонній порівняльний аналіз верхніх критичних магнітних полів $\mu_0 H_{c2}(0)$ в рамках теорій Гінзбурга–Ландау (ГЛ) і Вертхамера–Гельфанда–Хоенберга (ВГХ) для оптимально допованих тонких плівок $YBa_2Cu_3O_{7-\delta}$. Для різних орієнтацій магнітного поля наші розрахунки дають 638 та 153 Тл у випадках $\mu_0 H_{c2}(0)$, $H \parallel ab$ та $\mu_0 H_{c2}(0)$, $H \parallel c$, відповідно, при $H_{c2}(0)$. Вперше визначено температурні залежності довжин когерентності $\xi_{ab}(T)$ та $\xi_c(T)$ відповідно запропонованим теоріям з використанням критеріїв 50 та 90% величини питомого опору в нормальному стані $\rho_N$. Підхід ГЛ (з критерієм $0{,}9\rho_N$) дозволяє отримати $\xi_{ab}(0) = 11{,}8$ Å та $\xi_c(0) = 3{,}0$ Å, що добре узгоджується з літературними даними. Обговорюються застосування дуже малих довжин когерентності у ВТНП.

Ключові слова: високотемпературні надпровідники, плівки YBCO, верхнє критичне поле, магнітне поле, довжина когерентності.